\documentstyle[sprocl,psfig]{article}
\def\Journal#1#2#3#4{{#1} {\bf #2}, #3 (#4)}
\def\NPA{{\em Nucl. Phys.} A}

\def\PLB{{\em Phys. Lett.}  B}
\def\PRL{\em Phys. Rev. Lett.}
\def\PRC{{\em Phys. Rev.} C}
\def\PRD{{\em Phys. Rev.} D}
\def\ZPA{{\em Z. Phys.} A}

\def\be{\begin{equation}}
\def\ee{\end{equation}}
\def\bea{\begin{eqnarray}}
\def\eea{\end{eqnarray}}
\begin{document}

\title{POLARIZATION RESPONSE FUNCTIONS IN KAON ELECTROPRODUCTION}

\author{ C. BENNHOLD }

\address{Center for Nuclear Studies, Department of Physics\\
The George Washington University\\
Washington, D. C. 20052, USA}

\author{ T. MART, D. KUSNO }

\address{Jurusan Fisika, FMIPA, Universitas Indonesia,
         Depok 16424, Indonesia}

\maketitle

\abstracts{The electroproduction of kaons on the nucleon is shown to be an 
ideal tool to extract electromagnetic form factors of strange mesons and
hyperons.  Longitudinal $n(e,e^{\prime}K^0)\Lambda$ cross sections are 
found to be sensitive to the $K^0$ form factor, while sensitivity to the 
$\Lambda$ magnetic form factor can be seen in recoil-beam double 
polarization observables.  The $K K^* \gamma$ transition form factor 
could be extracted by measuring transverse response functions in the 
$K \Sigma$ channels.}

\section{Structure Functions in Kaon Electroproduction}

The recent advancement of TJNAF and other high-duty cycle, continuous beam 
electron accelerators has sparked increased interest in the form factors 
of baryons and mesons. Experimental programs have been completed
at TJNAF to measure the form factor of the  charged pion and the charged 
kaon, while a number of other approved experiments aim at precise measurements 
of the proton and neutron form factors. On the theoretical side, the recent 
advances on improved actions in Lattice Gauge calculations should soon lead 
to reliable results for strange hadron form factors. However, except for the 
$K^+$ no  calculation up to now has given a quantitative prediction of the 
effect of these form factors on experimental observables accessible at 
these facilities.

\begin{table}[!ht]
\caption{Complete response functions for the pseudoscalar meson
electroproduction~\protect\cite{knoechlein}.
         The polarization of the target (recoil) is indicated by
         $\alpha$ ($\beta$). The last three columns ($^cTL'$, $^sTL'$, 
         and $TT'$) are response functions for the polarized electron.
         $\ddagger$ denotes a response function which does not vanish but is
         identical to another response function.}
\renewcommand{\arraystretch}{1.4}
\begin{center}
\begin{tabular}{|cc|cccccc|ccc|}
\hline\hline
$\beta$ & $\alpha$ & $T$ & $L$ & $^cTL$ & $^sTL$ & $^cTT$ & 
 $^sTT$ & $^cTL'$ & $^sTL'$ & $TT'$\\
\hline\hline
-&-& $R_T^{00}$ & $R_L^{00}$ & $R_{TL}^{00}$ & 0 & $R_{TT}^{00}$ & 
 0 &  0 & $R_{TL'}^{00}$ & 0 \\
\hline
-&$x$ & 0 & 0 & 0 & $R_{TL}^{0x}$ & 0 &  $R_{TT}^{0x}$ & $R_{TL'}^{0x}$ &
 0 &  $R_{TT'}^{0x}$ \\
-&$y$ & $R_T^{0y}$ & $R_L^{0y}$ & $R_{TL}^{0y}$ & 0 & $\ddagger$ & 0 & 0 & 
  $R_{TL'}^{0y}$ & 0 \\
-&$z$& 0 & 0 & 0 & $R_{TL}^{0z}$ & 0 & $R_{TT}^{0z}$ & $R_{TL'}^{0z}$ & 0 & 
  $R_{TT'}^{0z}$ \\
\hline
~$x'$~&-& 0 & 0 & 0 & $R_{TL}^{x'0}$ & 0 &  $R_{TT}^{x'0}$ & $R_{TL'}^{x'0}$ &
 0 &  $R_{TT'}^{x'0}$ \\
$y'$&-& $R_T^{y'0}$ & $\ddagger$ & $\ddagger$ & 0 & $\ddagger$ & 0 & 0 & 
$\ddagger$ & 0 \\
$z'$&-& 0 & 0 & 0 & $R_{TL}^{z'0}$ & 0 & $R_{TT}^{z'0}$ & $R_{TL'}^{z'0}$ &0& 
  $R_{TT'}^{z'0}$ \\
\hline
$x'$&$x$& $R_T^{x'x}$ & $R_L^{x'x}$ & $R_{TL}^{x'x}$ & 0 & $\ddagger$ & 
 0 &  0 & $R_{TL'}^{x'x}$ & 0 \\
$x'$&$y$&0&0&0&$\ddagger$&0&$\ddagger$&$\ddagger$&0&$\ddagger$\\
$x'$&$z$& $R_T^{x'z}$ & $R_L^{x'z}$ &$\ddagger$&0&$\ddagger$&0&0&$\ddagger$&0\\
$y'$&$x$&0&0&0&$\ddagger$&0&$\ddagger$&$\ddagger$&0&$\ddagger$\\
$y'$&$y$&$\ddagger$&$\ddagger$&$\ddagger$&0&$\ddagger$&0&0&$\ddagger$&0\\
$y'$&$z$&0&0&0&$\ddagger$&0&$\ddagger$&$\ddagger$&0&$\ddagger$\\
$z'$&$x$& $R_T^{z'x}$ & $\ddagger$ & $R_{TL}^{z'x}$ & 0 & $\ddagger$ & 
 0 &  0 & $R_{TL'}^{z'x}$ & 0 \\
$z'$&$y$&0&0&0&$\ddagger$&0&$\ddagger$&$\ddagger$&0&$\ddagger$\\
$z'$&$z$&$R_T^{z'z}$&$\ddagger$&$\ddagger$&0&$\ddagger$&0&0&$\ddagger$&0\\
\hline\hline
\end{tabular}
\end{center}
\end{table}

In this paper we study the sensitivity of various kaon electroproduction
reactions to the electromagnetic form factors of the involved hyperons
and mesons.  Allowing for polarization in beam, target and recoil, the
differential cross section for kaon production with a virtual photon
can be written as~\cite{knoechlein}
\begin{eqnarray}
\lefteqn{
\frac{d\sigma_v}{d\Omega_K} ~=~ \frac{\mid {\vec q}_K \mid}{k_{\gamma}^{cm}}
P_{\alpha} P_{\beta} \left\{ R_T^{\beta \alpha}
+ \varepsilon_L R_L^{\beta \alpha } + \left[ 2 \varepsilon_L \left( 1 +
\varepsilon \right) \right]^{\frac{1}{2}} \right. \times } \nonumber \\
& & ( ^c \! R_{TL}^{\beta \alpha}
\cos \phi_{K} + ^s \! \! R_{TL}^{\beta \alpha} \sin \phi_{K} ) 
+ \varepsilon ( ^c \! R_{TT}^{\beta \alpha} \cos 2 \phi_{K}
+ ^s \! \! R_{TT}^{\beta \alpha} \sin 2 \phi_{K} ) + \nonumber \\
& & h \left[ 2 \varepsilon_L ( 1 - \varepsilon ) \right]^{\frac{1}{2}}
\left. ( ^c \! R_{TL'}^{\beta \alpha} \cos \phi_{K}
+ ^s \! \! R_{TL'}^{\beta \alpha} \sin \phi_{K} ) 
+ h (1-\varepsilon^2 )^{\frac{1}{2}} R_{TT'}^{\beta \alpha} \right\}, ~~
\end{eqnarray}
where $P_{\alpha} = (1, \vec P)$ and $P_{\beta} = (1, \vec P')$.
Here $\vec P = (P_x, P_y, P_z)$ denotes the target
and $\vec P' = (P_{x'}, P_{y'}, P_{z'})$ the recoil polarization vector.
Due to the self-analyzing property of the hyperon in the final state
the kaon electroproduction process is uniquely suited to measure
multiple polarization response functions.  In total, there are 36 different 
response functions in any pseudoscalar meson electroproduction 
experiment~\cite{knoechlein}, listed in Table 1. With a polarized beam 
already available at TJNAF all of these observables can be 
accessed once a polarized target is in place.   This opens the door
for a complete experiment in kaon electroproduction with much less
experimental effort than would be required for pion and eta 
electroproduction.

Over the last several years considerable effort has been spent to
develop models for the electromagnetic production of kaons from
nucleons at low energies~\cite{abw,saghai96,as1,cota,mart}. 
Due to the limited set of data the various models permit only qualitative
conclusions but do not yet allow the extraction of precise coupling
constants and resonance parameters. We employ the kaon electroproduction 
model of Ref.~\cite{cota} along with the methods of Ref.~\cite{mart} to 
extend this model to the $(e,e^{\prime }K^0)$ channels. 
Since at present no data are available for the $(e,e^{\prime }K^{0}) $
process it is impossible to test the reliability of the above amplitude
in this production channel. Therefore, rather than making precise
quantitative predictions for counting rates we emphasize here
 the relative  effects that the form factors have on
particular response functions.

\section{The $K^0$ Form Factor}

Among the SU(3) pseudoscalar mesons, the neutral kaon is the only
neutral system that can have a nonvanishing electromagnetic form
factor at finite $q^2$. Invariance of the strong interaction under 
charge conjugation dictates that the form factors of antiparticles 
are just the negative of those of the respective particles, thus, 
the $\pi ^0$ and $\eta $ do not have any electromagnetic form factors. 
Although the strange quark is notably heavier than the up and down
quarks, the mass difference is still smaller than the mass scale
associated with confinement in Quantum Chromodynamics (QCD),
$(m_s-m_d)<\Lambda _{\rm QCD}$. Therefore, if this mass difference 
affects physical observables, it could lead to a sensitive test of 
phenomenological models that attempt to describe nonperturbative QCD.

\begin{figure}[!ht]
\begin{minipage}[t]{57.5mm}
{\psfig{figure=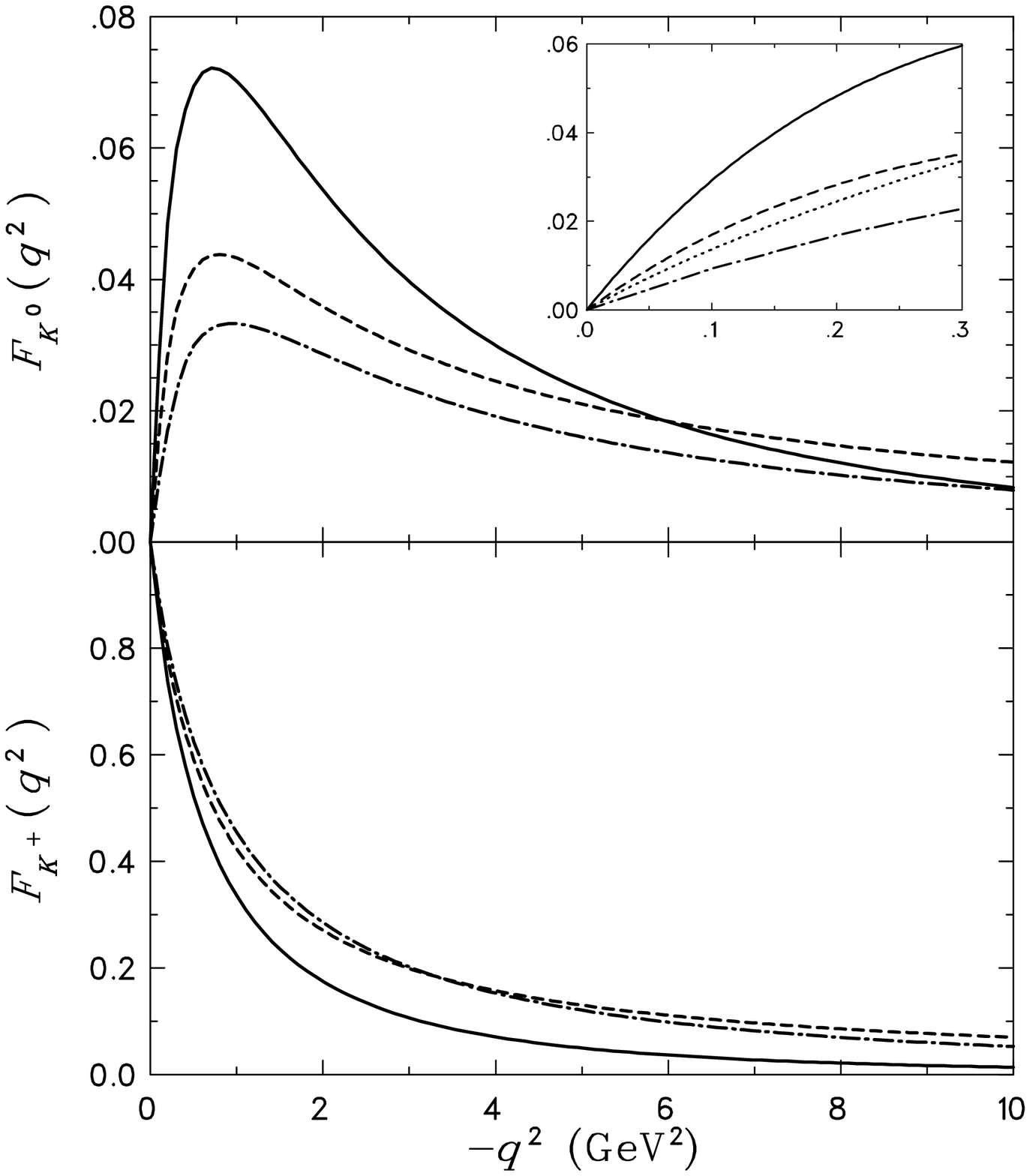,width=5.75cm}}
\caption{The $K^0$ and $K^+$ form factors calculated in the different
models. The dash-dotted line represents the QMV calculation, the
dashed line shows the VMD model, the dotted line shows the result of
$\chi$PT, while the solid line comes from LCQ model. The units of the
insert are the same as the one for the larger figure.}
\label{fig:ffk0}
\end{minipage}
\hspace{\fill}
\begin{minipage}[t]{57.5mm}
{\psfig{figure=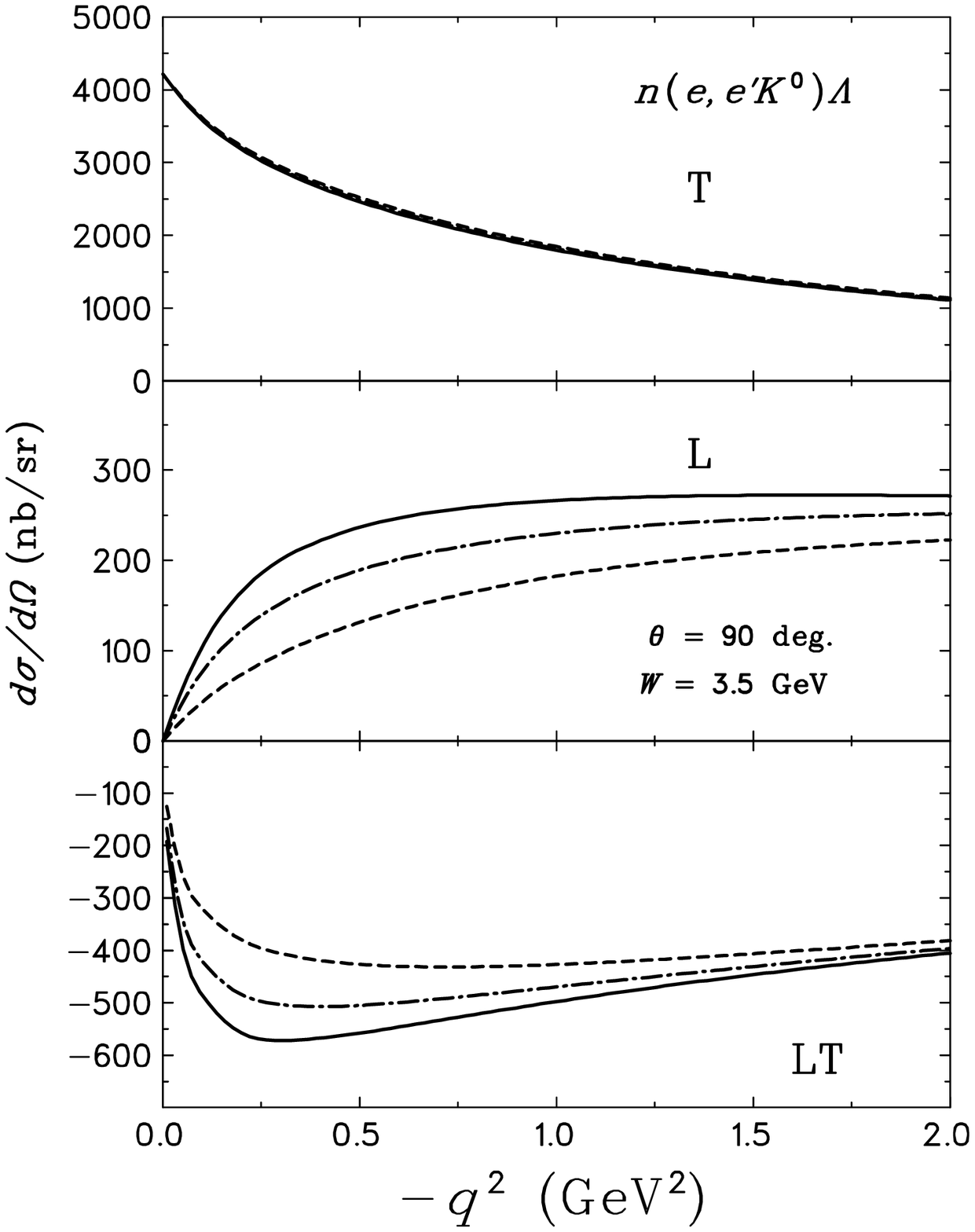,width=5.75cm}}
\caption{Transverse, longitudinal and longitudinal-transverse cross 
sections for the $n(e,e^{\prime}K^0)\Lambda$ process. The solid line 
shows the calculation with a $K^0$ form factor obtained in the LCQ model
while the dash-dotted line was obtained using the QMV model.
The dashed line shows a computation with the $K^0$ pole excluded.}
\label{fig:csffk0}
\end{minipage}
\end{figure}

We employ two relativistic quark models to calculate the $K^0$ form 
factor,  the light-cone quark (LCQ) model~\cite{ito1} and the quark-meson
vertex (QMV) model~\cite{buck}; 
several recent studies have put forth 
additional new model calculations of the $K^0$ electromagnetic form
factor~\cite{muenz96,salme96}. The charge form factor of the $K^0$ meson 
can be expressed as 
\begin{eqnarray}
  F_{K^0}(q) &=& e_d F_L(q) + e_{\bar s}F_H(q) ~, 
\label{eq:fk0}
\end{eqnarray}
where $F_L(q)$ and $F_H(q)$ are two independent form factors generated by 
the interaction of the photon with the light quark ($d$) of charge 
$e_d$ and   with the other heavier quark (${\bar s}$) of charge $e_{\bar s}$.
The results are compared with predictions
from vector meson dominance (VMD) and chiral perturbation
theory ($\chi$PT). In VMD, 
we assume that the photon interacts with the
strange quark through the $\phi $-meson and with the $u$-and
$d$- quarks through $\rho $- and $\omega $-mesons, each of them
having the strength proportional to the   quark charge. By ignoring
the mass difference of the $\rho $- and $\omega $%
-mesons, we can construct a simple two-pole model of the kaon form factors
via $F_L(q) = m_\omega ^2 / (m_\omega ^2-q^2)$ and 
$F_H(q) = m_\phi ^2 / (m_\phi ^2-q^2)$.
Using $\chi$PT to order $p^4$, a parameter-free
prediction for the $K^0$ form factor at very low $q^2$ 
can be obtained which is due 
entirely to one-loop diagrams without a tree-level contribution.
All models can reproduce the $K^+$ radius well; the experimental 
uncertainty of the $K^0$ radius does not yet distinguish between the different
models.

Figure~\ref{fig:ffk0} shows the form factors of the neutral kaon calculated 
in these various models. The charged kaon form factor is shown for comparison.
Note that the prediction of $\chi$PT is only valid for $-q^2 < 
0.3~{\rm GeV}^2$. The $K^0$ form factor has a peak around $-q^2\simeq 
1~{\rm GeV}^2$ for the other three models, though the heights are different. 
This peak is due to the behavior of the two different form factors, 
$F_L(q)$ and $F_H(q)$, that define $F_{K^0}$ as shown in Eq.~(\ref{eq:fk0}). 
$F_H(q)$ is generated by the coupling of the photon to the heavy quark and, 
therefore, falls off slower than $F_L(q)$ which comes from the light quark. 
Once the form factors are weighted by the quark charges, 
$e_d = -\frac{1}{3}$ and $e_{\bar s} = +\frac{1}{3}$, the general 
shape of $F_{K^0}$ is given.

In Fig.~\ref{fig:csffk0} we present predictions for the longitudinal and
transverse differential cross section of the $n(e,e^{\prime}K^0)\Lambda$ 
reaction using two different quark models for the $K^{0}$ form factor. 
While the transverse response is insensitive to the $K^0$ pole
the longitudinal cross section computed
with the form factor obtained from the LCQ-model is almost 50$\%$ larger
than the calculation with no $K^{0}$
pole while the QMV-model calculation lies between those two.
The similar sensitivities of the L and LT cross
sections indicate that a Rosenbluth separation is not imperative in
order to isolate the $K^{0}$ form factor effects. 

\section{The $\Lambda$ Form Factor}

\begin{figure}[!ht]
\centerline{\psfig{figure=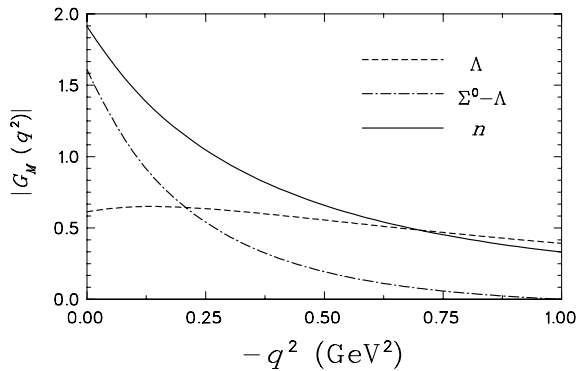,width=7.5cm}}
\caption{The $\Lambda$ and $\Lambda - \Sigma^0$ 
transition magnetic form factors as predicted by 
Ref.~\protect\cite{williams97} compared with that of the neutron.}
\label{fig:ffla}
\end{figure}

Interest in the hyperon electromagnetic form factors is related to the 
question of SU(3) flavor symmetry breaking and the effects of explicit 
and hidden strangeness in electromagnetic observables.  Applying SU(3) 
symmetry  allows  predicting the hyperon form factors in vector meson
dominance, quark and soliton models in terms of model parameters fixed 
by nucleon data. However, there is no direct method of measuring hyperon 
form factors since there are no stable hyperon targets. Here we suggest 
that kaon electroproduction may provide at least an indirect method  of 
obtaining information on these form factors.  Since the hyperon appears
in the $u$-channel one would want to choose kinematics with small $u$ and
large $t$ to suppress $t$-channel contributions.   

\begin{figure}[!ht]
\begin{minipage}[t]{57.5mm}
\centerline{\psfig{figure=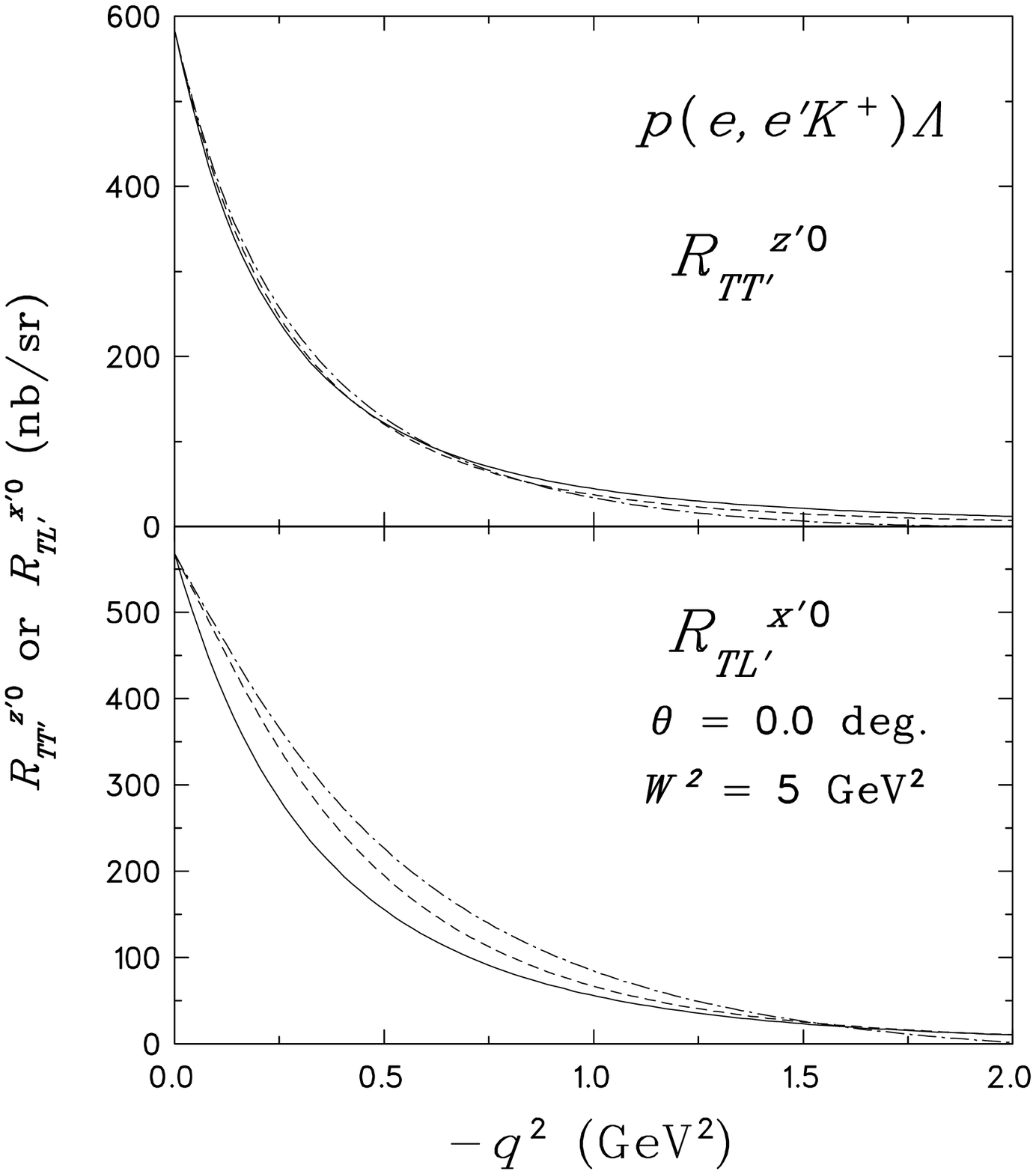,width=5.75cm}}
\caption{The sensitivity of response functions to different models of the
 $\Lambda$ form factor. The solid line corresponds to the neutron magnetic 
 form factor, the dash-dotted line is obtained by using a point-particle 
 approximation, while the dashed line is due to the $\Lambda$ form factor
 given by Ref.~\protect\cite{williams97}.}
\label{fig:rttp1}
\end{minipage}
\hspace{\fill}
\begin{minipage}[t]{57.5mm}
\centerline{\psfig{figure=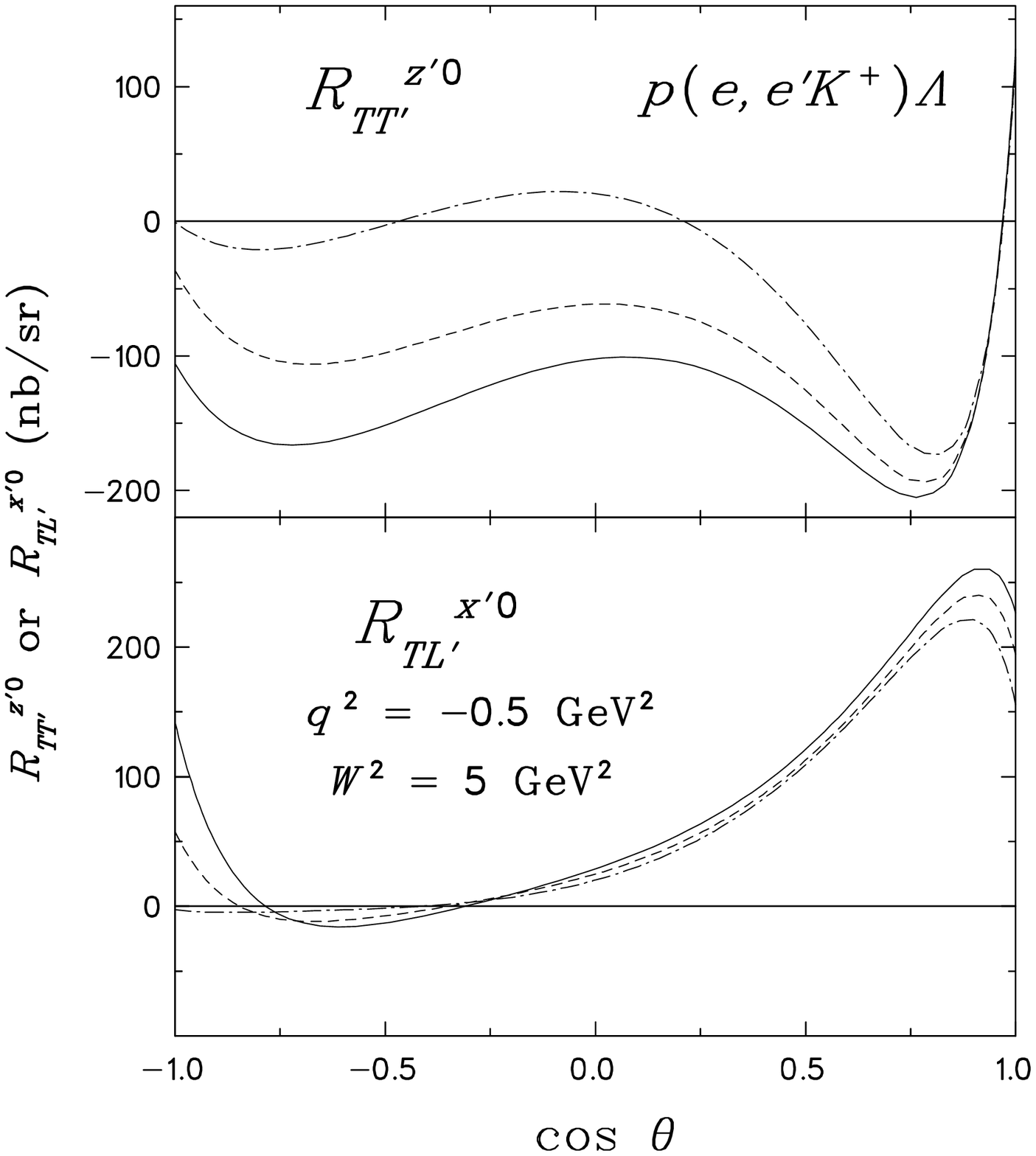,width=5.75cm}}
\caption{As in Fig.~\ref{fig:rttp1}, but for different kinematics.}
\label{fig:rttp2}
\end{minipage}
\end{figure}

Figure \ref{fig:ffla} shows the $\Lambda$ magnetic form factor as predicted in 
Ref.~\cite{williams97}.  They used a hybrid vector meson dominance (VMD)
formalism which provides a smooth transition from the low-$q^2$
behavior predicted by vector meson dominance to the high-$q^2$
scaling of perturbative QCD.  The key feature is the application
of the universality limit of the vector meson hadronic coupling SU(3) 
symmetry relations. 
Using a direct photon coupling along with a
$\phi$ and $\omega$ pole they predict the $\Lambda$ form factor,
while the $\Lambda - \Sigma$ transition form factor 
is obtained similarly, using a direct photon coupling and a $\rho$ pole.
Theoretically, the ratio of these form factors would be interesting since
the $\Lambda$ form factor depends only on isoscalar currents
while the $\Lambda - \Sigma$ transition depends only on isovector
contributions. Hence, as pointed out in Ref.~\cite{williams97}, this ratio
might see explicit strangeness and OZI effects such as
the suppression or enhancement of effective $\rho$, $\omega$ and $\phi$
vector meson-hyperon couplings relative to the vector meson-nucleon
couplings and SU(3) flavor symmetry predictions.
In comparison, we show the neutron magnetic form factor with a 
simple, standard
dipole parametrization.  Most previous $(e,e'K)$ studies have used
this neutron form factor parametrization
for the $\Lambda$, scaled by the magnetic moment.
Clearly, the $q^2$-dependence of the $\Lambda$ form factor predicted
by VMD is very
different from the shape of the neutron form factor with a much slower 
fall-off.

Figures \ref{fig:rttp1} and \ref{fig:rttp2} display double polarization 
response functions for the $p(e,e' K^+) \Lambda$ reaction that involve beam
as well as recoil polarization. These observables were subject of a recent
TJNAF proposal~\cite{baker97}.  
While the $T L'$ response shows moderate sensitivity to the different
$\Lambda$ form factors at very forward and backward angles, the $T T'$
structure function displays 
large sensitivities  over
almost the entire angular range, changing
both the magnitude and the shape of the cross section.  
Measuring these response functions could be accomplished
with CLAS in TJNAF's Hall B.

\section{The $K K^* \gamma$ Transition Form Factor}

\begin{figure}[!ht]
\centerline{\psfig{figure=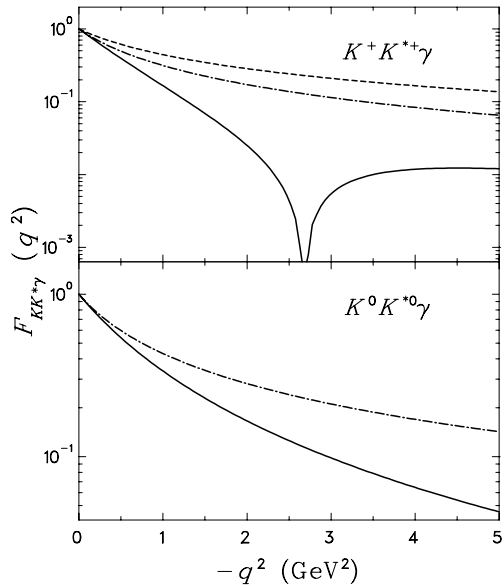,width=6.5cm}}
\caption{The transition $K^+K^{*+}\gamma$ and $K^0K^{*0}\gamma$
form factors as predicted by Ref.~\protect\cite{cota} (dash-dotted lines)
and Ref.~\protect\cite{muenz96} (solid lines).}
\label{fig:fftr}
\end{figure}

\begin{figure}[!ht]
\begin{minipage}[t]{57.5mm}
\centerline{\psfig{figure=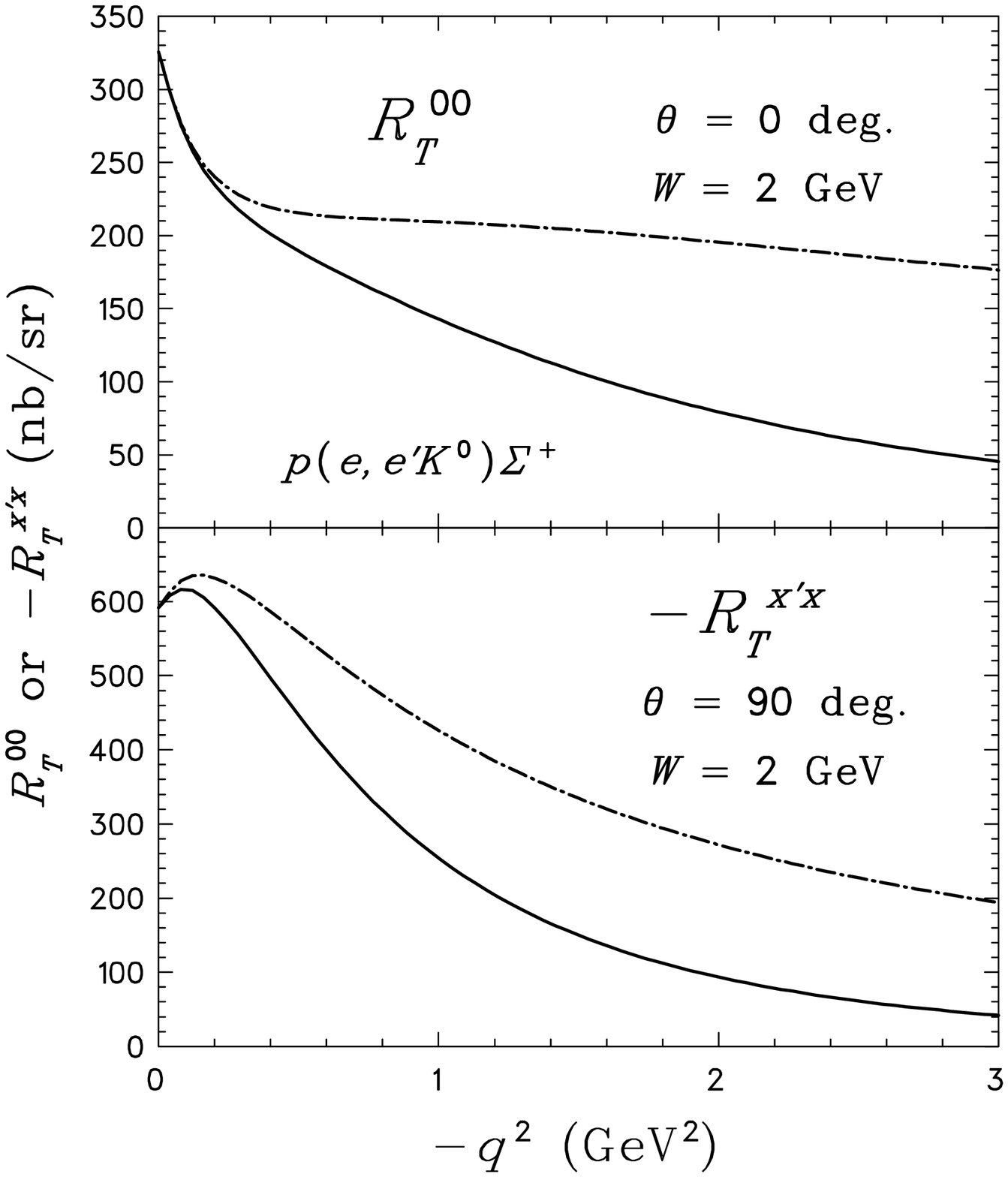,width=5.75cm}}
\caption{The sensitivity of response functions to different $K^0K^{*0}\gamma$
 form factors. The solid (dash-dotted) line is obtained by using the model 
 of Ref.~\protect\cite{muenz96} ( \protect\cite{cota}).}
\label{fig:k0tr}
\end{minipage}
\hspace{\fill}
\begin{minipage}[t]{57.5mm}
\centerline{\psfig{figure=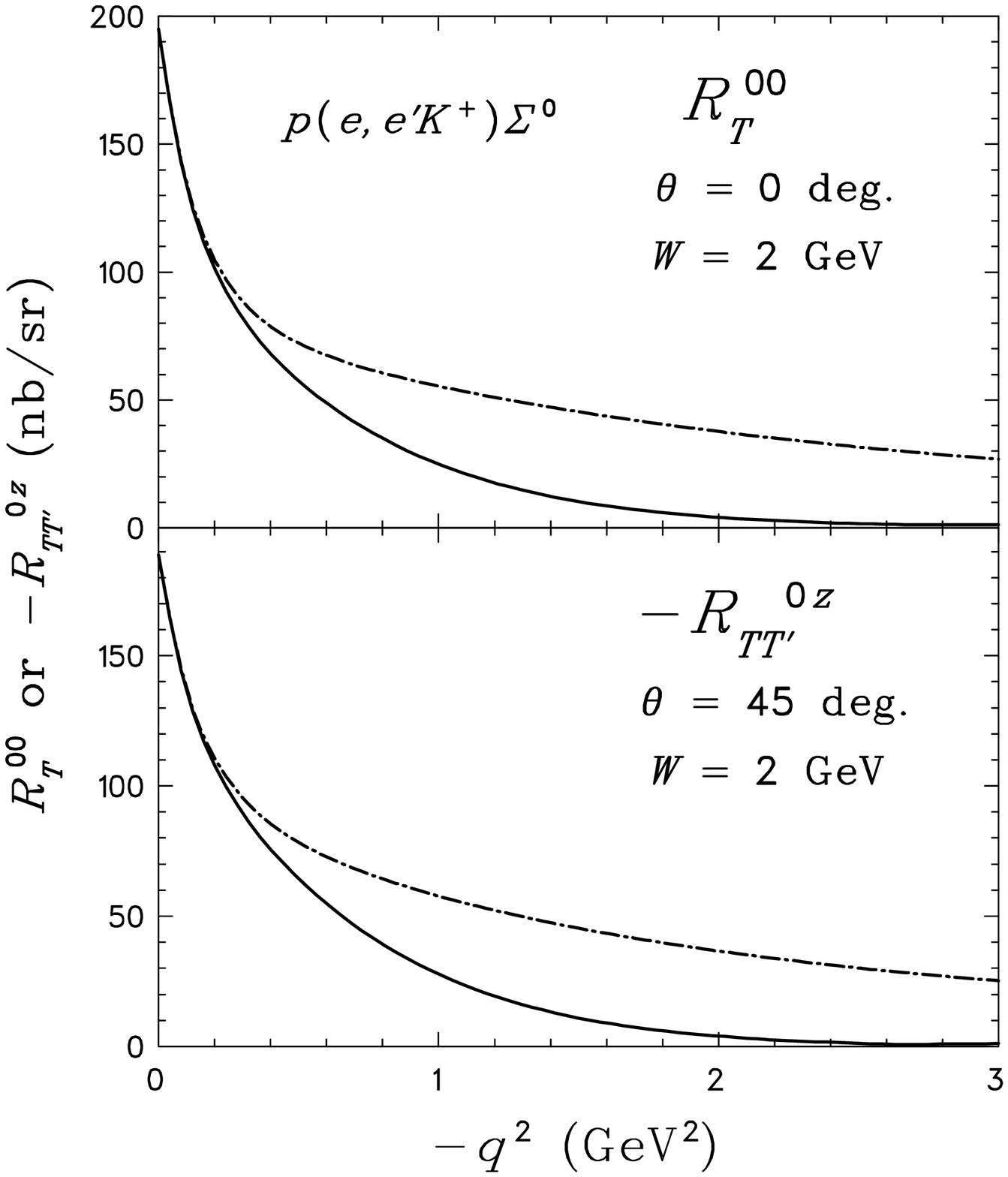,width=5.75cm}}
\caption{As in Fig.~\ref{fig:k0tr}, but for the $K^+K^{*+}\gamma$ form 
 factors.}
\label{fig:kptr}
\end{minipage}
\end{figure}

The $K K^* \gamma$ form factors have not been studied as
intensely as their non-strange counterparts, the $\rho \pi \gamma$
and $\omega \pi \gamma$ form factors which are very important in
meson exchange current corrections to deuteron electrodisintegration.
For the $K K^* \gamma$ transition
the form factors are again very sensitive to the mass difference 
between strange and non-strange quarks.  Figure \ref{fig:fftr} 
shows the transition form factors for both the neutral and the 
charged case, comparing the model of Ref.~\cite{cota} which uses vector 
meson dominance and the calculation of Ref.~\cite{muenz96} which solves a 
covariant Salpeter equation for a confining plus instanton-induced interaction.
Both models fall off faster than the elastic $K^+$ form factor which is
shown for comparison. The form factor in the charged case displays
a zero at higher $q^2$,
 indicating destructive interference between the light and the
heavy quark contribution.

Figures \ref{fig:k0tr} and \ref{fig:kptr} display the sensitivity of 
different transverse response
functions to the transition form factors. Since the small size of 
the $g_{K \Sigma N}$ coupling constant suppresses the Born terms
the $p(e,e' K^+)\Sigma^0$ reaction is used to study the $K^+ K^{*+}
\gamma$ form factor while the $p(e,e' K^0)\Sigma^+$ is sensitive
to the neutral transition.  Questions remain regarding additional
$t$-channel resonance contributions from states like the $K_1$(1270) 
which would have a different transition form factor. 
The observables displayed can clearly
distinguish between the different models, with the model of
Ref.~\cite{muenz96} leading to a much faster fall-off. 
Unfortunately, the zero in the charged transition form factor
around $q^2 = -2.6$ GeV$^2$ is not visible since the cross sections
are already very small at these momentum transfers.

\section{Conclusion}

In this paper we have shown that the kaon 
electroproduction process is well suited
to extract the form factors of strange mesons and baryons.
However, this cannot be done
 model-independently like a Chew-Low extrapolation
in the case of the charged pion or kaon.
In order to  reduce the model dependency and obtain accurate
quantitative predictions one requires the information from
the TJNAF kaon photoproduction experiments~\cite{schu}. The
analysis of those measurements should determine the relevant
resonances and coupling constants and thus uniquely define the
production amplitude.

\section*{Acknowledgments}

We are grateful to K. Dhuga and O.K. Baker for useful discussions 
regarding the experimental aspects. This work is supported by the 
US DOE grant no. DE-FG02-95-ER40907 and the University Research for 
Graduate Education (URGE) grant.

\end{document}